\documentstyle[preprint,pra,aps]{revtex}

\begin{document}
\draft

\title{Conservation laws in the continuum $1/\lowercase{r}^2$ systems}

\author{Rudolf A.\ R\"{o}mer\cite{rar}}

\address{Condensed Matter Theory Unit,
Jawaharlal Nehru Centre for Advanced Scientific Research,
Indian Institute of Science Campus,
Bangalore, 560 012, India}

\author{B.\ Sriram Shastry}

\address{Physics Department, Indian Institute of Science, Bangalore 560 012,
India}


\author{Bill Sutherland}

\address{Physics Department, University of Utah, Salt Lake City, Utah 84112,
U.S.A.}

\date{Version: October 4, 1995; printed \today}
\maketitle

\begin{abstract}
We study the conservation laws of both the classical and the
quantummechanical continuum $1/r^2$ type systems.
For the classical case, we introduce new integrals of motion
along the recent ideas of Shastry and Sutherland (SS),
supplementing the usual integrals of motion constructed
much earlier by Moser.
We show by explicit construction that one set of integrals can
be related algebraically to the other. The difference of
these two sets of integrals then gives rise to yet another
complete set of integrals of motion.
For the quantum case, we first need to resum the integrals
proposed by Calogero, Marchioro and Ragnisco. We give a
diagrammatic construction scheme for these new integrals,
which are the quantum analogues of the classical traces.
Again we show that there is a relationship between these new
integrals and the quantum integrals of SS by explicit
construction.
Finally, we go to the asymptotic or low-density limit and
derive recursion relations of the two sets of asymptotic 
integrals.
\end{abstract}


\narrowtext
\tighten

%
%

\section{Introduction}

The integrability of both the classical and the quantum  
one-dimensional problem of $N$ particles interacting via the two-body 
potentials $V_0(x)= g^2/x^2$, $V_t(x)= g^2 \Phi^2 \sin^{-2}[ \Phi x ]$ 
and $V_h(x)= g^2 \Phi^2 \sinh^{-2}[ \Phi x ]$ has been shown more 
than two decades ago by Moser \cite{moser} (for the classical problem) 
and Calogero, Marchioro and Ragnisco \cite{cmr} (for the quantum 
problem), both groups exploiting a technique due to Lax 
\cite{lax}.
These early results have been reviewed, extended and collected 
nicely both for the classical and the quantum cases 
by Olshanetsky and Perelomov in Ref.\ \cite{opc,opq}.

For the classical systems, integrability restricts
the motion in terms of action-angle variables onto a torus in
phase space.
However, for the quantum case, integrability leads to solvability
only for those special cases which support scattering, i.e.,
systems which fly apart when the walls of the box are removed.
In these cases, integrability implies conservation of 
individual momenta and thus the wave function is given
asymptotically by Bethe's Ansatz. 
For the above interaction potentials, Sutherland \cite{bill}
has exploited this fact to determine the properties of the
quantum systems in the thermodynamic limit. 

Recently, Shastry and Sutherland \cite{ss} have given an
independent proof of integrability of the quantum many-body 
problem and constructed new integrals of motion. However, for any 
finite number of particles $N$, we know that in principle we have 
exactly $N$ 
conserved quantities. Therefore we expect the new integrals of motion
to be 
related to the integrals constructed by CMR. It is the aim of the
present work to elucidate some of the features of the new 
integrals of motion and to show their relation to the integrals of CMR.
We emphasize that this new proof of integrability has also made
possible the application of the ideas of the asymptotic Bethe
Ansatz to the $1/r^2$ models with quantum exchange \cite{srs}.

In section \ref{sec-classical} we show that the new construction 
of SS gives integrals of motion also for the classical problem.
We next explicitly calculate these new integrals up to $n=4$ and 
compare them to the integrals of CMR. This then gives rise 
to yet another set of integrals $K_n$. Section \ref{sec-quantum} is
devoted to a comparison of the two series of integrals of motion
 for the 
quantum case. The integrals given by CMR are not extensive
quantities and we need to resum them via an application of the 
linked cluster theorem.
In section \ref{sec-limit}, we take the asymptotic or  
low-density limit of the problem and section \ref{sec-concl}
summarizes and discusses our results.

%
%

\section{The classical case}
\label{sec-classical}

The Hamiltonian of primary interest for our present work is 
given as
\begin{equation}
H= \sum_{i} p_i^2 + 
        \lambda(\lambda-1) \Phi^2 \sum_{ij}' \sinh^{-2} [\Phi (x_i-x_j)].
\label{eq-hcmr}
\end{equation}
The interaction term reduces to $V_0$ in the limit of high-densities
(or $\Phi\rightarrow 0$) and the trigonometric interaction $V_t$ is just
the analytic continuation of $\Phi\rightarrow i\Phi$.
Here and in the following, we will use the primed sum $\sum'$ to indicate
that the summation runs over unequal indices only. 

%
%
\subsection{Moser's invariants}

Let us briefly recall the method of $\cite{moser,lax}$:
We introduce the Lax pair $L$, $M$,
\begin{eqnarray}
L_{jk}
 &= &
 p_j \delta_{jk} +
 i (1-\delta_{jk}) \sqrt{\lambda(\lambda-1)} \Phi \coth [\Phi(x_j-x_k) ], \\
M_{jk}
 &= &
 2\sqrt{\lambda(\lambda-1)} \Phi^2 \left[ \delta_{jk} 
 \sum_{l}' \sinh^{-2}[\Phi(x_j-x_l) ]
 + (1-\delta_{jk}) \sinh^{-2}[\Phi(x_j-x_k) ] \right].
\label{eq-lcmr}
\end{eqnarray}
The classical equations of motion then imply the matrix equation
\begin{equation}
 \frac{dL}{dt} = \{L,H\} = i [ML-LM],
\label{eq-lax}
\end{equation}
where we define the Poisson brackets as
$\{ F, G \} = \sum_{j=1}^{N} 
 \frac{\partial F}{\partial x_j} \frac{\partial G}{\partial p_j} 
  -
 \frac{\partial F}{\partial p_j} \frac{\partial G}{\partial x_j}
$. 
The time evolution of $L$ consequently is an isospectral 
deformation, 
$L(t)= \exp[ i \int_0^t M(\tau) d\tau ] L(0) \exp[-i\int_0^t M(\tau) d\tau]$,
and the integrals of motion are simply given as the traces 
\begin{equation}
 T_n = \mbox{Tr} L^n(t)= \mbox{Tr} L^n.
\end{equation}

We also need to show that the $T_n$'s are in involution, e.g., 
$\{ T_n, T_m \}=0$.
Using the Jacobi relation for Poisson brackets, we see that
\begin{equation}
\{ H, \{T_n,T_m\} \} = 
 \{ T_n, \{H,T_m\} \} - \{ T_m, \{H,T_n\}\},
\end{equation}
and thus $\{ T_n, T_m \}$ is also an integral of motion. 
But, allowing the system the evolve in time, all particles scatter,
the Lax matrix itself evolves into
\begin{equation}
L \stackrel{t\rightarrow\infty}{\longrightarrow}
L^{\infty} =
\left\{ \begin{array}{ll}
L_{jj} = k_j, & \\
L_{jk} = +i \sqrt{\lambda(\lambda-1)}, &j>k, \\
L_{jk} = -i \sqrt{\lambda(\lambda-1)}, &j<k.
\end{array} \right.
\label{eq-linfty}
\end{equation} and so the coordinate dependence vanishes. 
Thus the Poisson bracket $\{ T_n, T_m \}$ evaluates to
zero. 
We remark that it is this procedure that we use to prove involution
for all the integrals constructed in the following chapters.

Let us define 
$\alpha_{jk} = 
\sqrt{\lambda(\lambda-1)} \Phi \coth [\Phi(x_j-x_k) ]$ and
$\alpha_{jj}=0$. 
Then a
direct calculation of the integrals of motion up to $n=4$ gives
\begin{mathletters}
\label{eq-ct}
\begin{eqnarray}
T_1
&= & \sum_{i} p_i = P, 
\label{eq-ct1} \\
T_2
&= & 
\sum_i p_i^2 + \sum_{ij}' \alpha_{ij}^2, 
\label{eq-ct2} \\
T_3
&= & 
\sum_i p_i^3 + 3 \sum_{ij}' \alpha_{ij}^2 p_i, 
\label{eq-ct3} \\
T_4
&= & \sum_i p_i^4 + 2 \sum_{ij}' \alpha_{ij}^2 (p_j^2 + p_i p_j + p_i^2)
+ \mbox{Tr}\alpha^4. 
\label{eq-ct4}
\end{eqnarray}
\end{mathletters}
Using $\alpha_{ij}^2 = \Phi \sqrt{\lambda(\lambda-1)}
\left[ \sinh^{-2}[\Phi(x_j-x_k)] + 1 \right]$,
we see that 
$T_2 = H + \Phi^2 {\lambda(\lambda-1)} N(N-1)$.
Note that due to the antisymmetry $\alpha_{ij}= -\alpha_{ji}$,
only even powers of $\alpha$ --- and thus integer powers of
$\lambda$ --- will appear in all these expressions.

Let us now define the classical down-boost \cite{opc}
\begin{equation}
X= \sum_{j=1}^N x_j.
\label{eq-boost}
\end{equation}
We then find easily that
\begin{equation}
\{ X, T_n \} = n T_{n-1}.
\label{eq-tboost}
\end{equation}
Further, Jacobi's identity gives
\begin{equation}
\{ X, \{T_n,T_m\} \} = 
 (n-1) \{ T_{n-1}, T_m \} + (m-1) \{ T_{m-1}, T_n\}.
\end{equation} 
As a particular case, suppose $n=2$, so $n-1=1$ and
$T_{n-1}=P$. Then by translation invariance
$\{ P, T_n\}=0$, so we conclude that if
$\{H,T_n\}=0$, then $\{H,T_{n-1}\}=0$. 
In particular, $\{H,T_N\}=0$ implies that all $T_n$ are
integrals. 
Finally, we may construct all integrals of motion from
$T_N$ by repeatedly using the boost $X$ in the 
representation
\begin{equation}
X = \sum_{j=1}^{N} \frac{\partial}{\partial p_j}.
\label{eq-boostrep}
\end{equation}

%
%
\subsection{Shastry's invariants}

In Ref.\cite{ss}, Shastry and Sutherland provide a set of integrals
of motion for the quantum problem. However, mimicking their arguments,
we can straightforwardly construct integrals of motion for the classical
case, too. Let us introduce the singular matrix $\Delta_{jk}= 1$ for all
$i,j$ and the vector $\eta_j= 1$ for all $j$. Then we define integrals 
of motion s.t.
\begin{equation}
J_n = \mbox{Tr}[ L^n(t) \Delta ] 
= \eta^{\dagger} L^n(t) \eta
= \sum_{i_1,i_2,\ldots, i_{n+1}} 
 L_{i_1 i_2} L_{i_2 i_3} \cdots L_{i_{n-1} i_{n}} L_{i_n i_{n+1}} , 
\end{equation}
with the Lax matrix $L$ given as before.
We then have
\begin{eqnarray}
\frac{d J_n}{dt}
 &= & \frac{d}{dt}\left\{ \mbox{Tr}\left[\exp[iMt] L^n(t) \exp[-iMt] \Delta \right] \right\} \\
 &= & i \mbox{Tr} [ M L^n(t) \Delta  -  L^n(t) M \Delta ] \\
 &= & i \left[ \mbox{Tr}[ L^n(t) \Delta M ] - \mbox{Tr} [ L^n(t) M \Delta ] \right]\\
 &= & 0,
\end{eqnarray}
since $M\Delta=\Delta M = 0$ as shown in SS.
Involution for these integrals of motion is proven by the same 
asymptotic argument as before. 
A direct calculation of the conserved quantities of SS up to $n=4$ 
gives:
\begin{mathletters}
\label{eq-cj}
\begin{eqnarray}
J_1 &= & \sum_{i} p_i, 
\label{eq-cj1} \\
J_2 &= & \sum_i p_i^2 + \sum_{ij}' \alpha_{ij}^2
        - \sum_{ijk}' \alpha_{ij} \alpha_{jk}, 
\label{eq-cj2} \\
J_3 &= & \sum_i p_i^3 + 3 \sum_{ij}' \alpha_{ij}^2 p_i
        -\sum_{ijk}' \alpha_{ij} \alpha_{jk} (p_i + p_j + p_k),
\label{eq-cj3} \\
J_4 &= &
\sum_i p_i^4
+2\sum_{ij}' \alpha_{ij}^2 \left[ p_i^2 + p_i p_j + p_j^2 \right]
+ \mbox{Tr} \alpha^4 
+ \sum_{i\neq j\neq k\neq l\neq m \neq i}
 \alpha_{ij} \alpha_{jk} \alpha_{kl} \alpha_{lm} \nonumber \\ & &
-\sum_{ijk}' \alpha_{ij} \alpha_{jk} \left[
p_i^2 + p_j^2 + p_k^2 + p_i p_j + p_j p_k + p_k p_i
\right].
\label{eq-cj4}
\end{eqnarray}
\end{mathletters}
Again the Hamiltonian can be found in the $n=2$ term,
$J_2= H + \Phi^2 {\lambda(\lambda-1)} N(N^2-1)/3$
and again only even powers of $\alpha$ appear in the 
expressions of the $J_n$'s.

The action of the down-boost on these new integrals of motion 
is as in Eq.\ (\ref{eq-tboost}), e.g., $\{ X, J_n \} = n 
J_{n-1}$. Much more useful is the up-boost $Y$ which we define 
as 
\begin{equation} 
 Y= \sum_i x_i p_i^2 + \sum_{ij}' (x_i + x_j) \alpha_{ij}^2 /2 
 \label{eq-upboostc} 
\end{equation} 
in analogy with the up-boost operator $\sum_n n {\bf S}_n {\bf 
S}_{n+1}$ in the Heisenberg model. Unfortunately, this up-boost 
only works, if we restrict ourselves to the potential $V_0$ 
such that $\alpha_{ij}^2= \lambda(\lambda-1) / ( x_i - x_j 
)^2$. In this case, we find by explicit construction that $\{Y, 
J_n\}= (n+1) J_{n+1}$. The Jacobi relation $\{ J_m, \{Y,J_n\} \} 
= \{ Y, \{J_m,J_n\}\} - \{ J_n, \{J_m,Y\}\}$ now gives 
\begin{equation} 
(n+1) \{ J_m, J_{n+1} \} =  
 \{ Y, \{J_m,J_n\}\} - (m+1) \{ J_n, J_{m+1} \}. 
 \label{eq-ubjacobic} 
\end{equation} 
Thus, if $\{J_m,J_n\}=0$ and $\{J_{m+1},J_n\}=0$, we also have 
$\{ J_m, J_{n+1} \} = 0$. We emphasize that the up-boost 
(\ref{eq-upboostc}) seems to work only for the special 
potential $V_0$.

%
%
\subsection{Relation between invariants}
\label{sec-crbi}

We can again use the Jacobi relation to show that the
Poisson bracket $\{ T_n, J_m \}$ is an integral of motion which
in the asymptotic limits evaluates to zero.
The difference between the integrals of motion of Moser and
SS then gives rise to yet another set of constants,
\begin{equation}
K_n = J_n - T_n = \sum_{i_1 \neq i_{n+1}} 
L_{i_1 i_2} L_{i_2 i_3} \cdots L_{i_{n-1} i_{n}} L_{i_n i_{n+1}}. 
\label{eq-kn}
\end{equation}
Various terms in the $J_n$'s can be simplified with the help of
the $\coth$-rule,
\begin{equation}
\alpha_{ij} \alpha_{jk} +
\alpha_{ij} \alpha_{ki} +
\alpha_{jk} \alpha_{ki} = -\Phi^2 \lambda(\lambda-1),
\label{eq-cothrule}
\end{equation}
and hence we find
\begin{mathletters}
\label{eq-ck}
\begin{eqnarray}
K_1 &= & 0, \\
K_2 &= & \Phi^2 \lambda(\lambda-1) N(N-1)(N-2)/3, \\
K_3 &= & \Phi^2 \lambda(\lambda-1) (N-1)(N-2) P, \\ 
K_4 &= & \Phi^2 \lambda(\lambda-1)
 (N-2) \left[ (N-2) T_2 + P^2 \right] +
\left[ \Phi^2 \lambda(\lambda-1) (N-1)(N-2) \right]^2/9. 
\end{eqnarray}
\end{mathletters}
Note that $K_3$ is the first term that is not a simple constant,
and in order to make the $K_n$'s a complete set of integrals
of motion, we may simply use $K_{N+1}$ and $K_{N+2}$.
Thus we conclude that by construction, we can express Shastry's
integrals of motion in terms of Moser's and vice versa.
We emphasize that this relationship is not linear, but only
algebraic as seen from the existence of the $P^2$ term in $K_4$.

Taking the limit $\Phi\rightarrow 0$, we see that the $K_n$'s 
are zero. Thus only for the simplest case of the Calogero potential
$V_0(x)= g^2/x^2$ do we find that the Moser set of integrals of 
motion is identical to the set of SS.

%
%

\section{The quantum case}
\label{sec-quantum}

In the quantum case, the elements of the Lax $L$ and $M$ matrices
become operators themselves, i.e., 
the momentum operator is $p_j= -i \partial/\partial x_j$
and we have the commutation relation
$[ x_j, p_k] = i \delta_{jk}$.
Since operator elements do not necessarily commute, we always mean 
an ordered product of elements when we multiply matrices in the 
following.

%
%
\subsection{Calogero's invariants}
\label{sec-qci}

The early work of Calogero, Marchioro and Ragnisco \cite{cmr}  
quantised the classical Lax equation, by antisymmetrizing the 
right-hand side of Eq.\ (\ref{eq-lax}).
The proof of invariance of the traces then does no longer hold. 
However, CMR also showed that after replacing the classical
variables with the corresponding quantum mechanical operators,
we can define new integrals of motion $I_n$ s.t.\
\begin{equation}
\Delta(\beta)\equiv\det[ 1-\beta L ] \equiv
 1 + \sum_{n=1}^{N} (-\beta)^n I_n.
\label{eq-in}
\end{equation} 
CMR then go on to argue that these $I_n$ are conserved, 
$[I_n,H]=0$, and in involution, $[I_n,I_m]=0$.
The later result is again proved \cite{asym} by use of the asymptotic 
limit as in the last section.
A direct calculation of the conserved quantities of CMR up to $n=5$ for
$H$ yields:
\begin{mathletters}
\label{eq-i}
\begin{eqnarray}
I_1 &= & \sum_{i} p_i, 
\label{eq-i1} \\
I_2 &= & \frac{1}{2} I_1^2 - 
        \frac{1}{2} \left[ \sum_i p_i^2 + \sum_{ij}' \alpha_{ij}^2 \right], 
\label{eq-i2} \\
I_3 &= & \frac{1}{6} \sum_{ijk}' p_i p_j p_k - 
        \frac{1}{2} \sum_{ijk}' \alpha_{jk}^2 p_i,
\label{eq-i3} \\
I_4
&= & \frac{1}{4!} \sum_{ijkl}' p_i p_j p_k p_l
        - \frac{1}{4} \sum_{ijkl}' \alpha_{ij}^2 p_k p_l 
\nonumber \\
&  & \mbox{ }
        -\frac{1}{4} \sum_{ijkl}' \alpha_{ij} \alpha_{jk} \alpha_{kl} \alpha_{li}
        + \frac{1}{8} \sum_{ijkl}' \alpha_{ij}^2 \alpha_{kl}^2,
\label{eq-i4} \\
I_5
&= & \frac{1}{5!} \sum_{ijklm}' p_i p_j p_k p_l p_m
        - \frac{1}{12} \sum_{ijklm}' \alpha_{ij}^2 p_k p_l p_m 
\nonumber \\
&  & \mbox{ }
        -\frac{1}{4} \sum_{ijklm}' \alpha_{ij} \alpha_{jk} \alpha_{kl} \alpha_{li} p_m
        + \frac{1}{8} \sum_{ijklm}' \alpha_{ij}^2 \alpha_{kl}^2 p_m.
\label{eq-i5}
\end{eqnarray}
\end{mathletters}
Note that the Hamiltonian can be found in the term in
parenthesis in $I_2$.

Let us define a quantum down-boost operator analogous to the classical
boost\cite{opq}. With $X= \sum_{j=1}^N x_j$ as before, we then find
\begin{equation}
[X,I_m] = i (N-m+1) I_{m-1}.
\end{equation}
Using Jacobi's identity for commutators, we can easily show
that as previously,
$[H,I_n]=0$ implies $[H,I_{n-1}]=0$ and thus
$[H,I_N]=0$ implies all $I_n$ are integrals.
A particularly nice result is to write
$I_N = \det L$, treat the momenta $p_j$ as classical
c-numbers since there are no ordering ambiguities, and
use the representation
\begin{equation}
X = \sum_{j=1}^{N} i \frac{\partial}{\partial p_j}
\end{equation}
to generate all $I_n$ in the quantum case.

Of special importance in the following will be that as in
the classical invariants by Moser, $\alpha$ will only appear
in even powers in the $I_n$'s. Therefore, $\lambda$ will 
occur with integer powers only and terms such as
$[ \lambda(\lambda-1) ]^{3/2}$ do not exist.

%
%
\subsection{Shastry's invariants}

In Ref.\ \cite{ss}, Shastry and Sutherland provide a proof of 
integrability in the quantum case via an entirely different method: 
The Hamiltonian $H$ is given as before but the
Lax matrices now read
\begin{eqnarray}
L^{SS}_{jk}
 &= &
 p_j \delta_{jk} +
 i (1-\delta_{jk}) \lambda \Phi \coth[\Phi(x_j-x_k)] \\
 & \equiv &
 p_j \delta_{jk} +
 i (1-\delta_{jk}) \chi_{jk}, \\
\label{eq-lss}
M^{SS}_{jk}
 &= &
 2\lambda \Phi^2 \left[ \delta_{jk} 
 \sum_{l}' \sinh^{-2}[\Phi(x_j-x_l) ]
 + (1-\delta_{jk}) \sinh^{-2}[\Phi(x_j-x_k) ] \right].
\label{eq-mss}
\end{eqnarray}
with $\chi_{ii}= 0$.
SS define their conserved quantum integrals of motion as 
in Eq.\ (\ref{eq-cj}), e.g., 
$J_n = \eta^{\dagger} (L^{SS})^n \eta$.
The new Lax matrices obey the ordered Lax equation
\begin{equation}
[ L^{SS}, H ] = M^{SS} L^{SS} - L^{SS} M^{SS},
\label{eq-olax}
\end{equation}
and we may easily prove invariance via
\begin{equation}
[ J_n, H ]
 = \eta^{\dagger} [ (L^{SS})^n, H ] \eta
 = \eta^{\dagger} \left[ M^{SS} (L^{SS})^n - (L^{SS})^n M^{SS} \right] \eta
 =0,
\end{equation}
since as before $\eta^{\dagger} M^{SS} = M^{SS} \eta = 0$.
A direct calculation of the conserved quantities of SS up to 
$n=4$ yields:
\begin{mathletters}
\label{eq-j}
\begin{eqnarray}
J_1 &= & \sum_{i} p_i, 
\label{eq-j1} \\
J_2 &= & \sum_i p_i^2 + \sum_{ij}' (\chi_{ij}^2+\chi'_{ij})
        - \sum_{ijk}' \chi_{ij} \chi_{jk}, 
\label{eq-j2} \\
J_3 &= & \sum_i p_i^3 + 3 \sum_{ij}' (\chi_{ij}^2+\chi'_{ij})p_i
        -\sum_{ijk}' \chi_{ij} \chi_{jk} (p_i + p_j + p_k),
\label{eq-j3} \\ 
J_4
&= &
\sum_i p_i^4
+2\sum_{ij}' (\chi_{ij}^2 + \chi_{ij}') 
   \left[ p_i^2 + p_i p_j + p_j^2 \right]
+ \sum_{i\neq j\neq k\neq l\neq m\neq i}
   \chi_{ij} \chi_{jk} \chi_{kl} \chi_{lm}
\nonumber \\ & &
+ \mbox{Tr} \chi^4
-\sum_{ijk}' \chi_{ij} \chi_{jk} \left[
p_i^2 + p_j^2 + p_k^2 + p_i p_j + p_j p_k + p_k p_i
\right] \label{eqn-j4cot}
\nonumber \\ & &
+ 2 i \sum_{ij}' \chi_{ij}'' p_j
+ 4 i \sum_{ij}' \chi_{ij} \chi_{ij}' p_j
+ i \sum_{ijk}' \chi_{ij} \chi_{jk}' (p_j - p_k)
\nonumber \\ & &
- \sum_{ij}' \chi_{ij}''' - 2 \sum_{ij} \chi_{ij} \chi_{ij}'' 
        + 2 \sum_{ijk}' \chi_{ij} \chi_{jk}''
\nonumber \\ & &
- \sum_{ij}' (\chi_{ij}')^2 + \sum_{ijk}'\chi_{ij}' \chi_{jk}'
+ 3 \sum_{ijk}' \chi_{ij}^2 \chi_{jk}' + 2 \sum_{ij}' \chi_{ij}^2 \chi_{ij}'
\nonumber \\ & &
- \sum_{ijkl}' \left[
 \chi_{ij} \chi_{jk}'\chi_{kl} + 2 \chi_{ij} \chi_{jk} \chi_{kl}'
\right]
+ \sum_{ijk}' \chi_{ij}\chi_{jk}\chi_{ki}'
\label{eq-j4}
\end{eqnarray}
\end{mathletters}
The derivative $\chi'_{jk}$ is defined by the commutator
$[ p_j, \chi^{(n)}_{jk} ] \equiv - i \chi^{(n+1)}_{jk} $.
See the appendix for an explicit list of derivatives.

Using 
$\chi_{ij}'= -\Phi^2 \lambda \sinh^{-2}[ \Phi (x_i - x_j) ]$,
we see that just as in the classical case, $J_2$ contains
the Hamiltonian, i.e., 
$J_2 = H + \Phi^2 \lambda^2 N (N^2-1)/3$.
However, the interaction strength $\lambda(\lambda-1)$
in the Hamiltonian could only be obtained with the modified
form of the Lax matrix $L^{SS}$. 
Also, the $\lambda$ dependence of the constant term in the above 
equation is different from its classical counterpart.
We remark that the last terms in Eq.\ (\ref{eq-j2}) and (\ref{eq-j3})
can again be written as $\mbox{const.}$ and 
$\mbox{const.}\times \sum_i p_i$ by the $\coth$-rule of Eq.\
(\ref{eq-cothrule}).

The down-boost operator acts as before, e.g., $[X,J_n]= i n J_{n-1}$.
In case of the potential $V_0$, we may also use the up-boost
of Eq.\ (\ref{eq-upboostc}) in operator form as
\begin{equation}
Y= \sum_i (x_i p_i^2 + p_i^2 x_i)/2 + \sum_{ij}' (x_i + x_j) \alpha_{ij}^2 /2.
\label{eq-upboost}
\end{equation}
Then $[Y, J_n]= i (n+1) J_{n+1}$ and we again have from the Jacobi 
identity
\begin{equation}
i (n+1) [ J_m, J_{n+1} ] =
 [ Y, [J_m,J_n] ] - i (m+1) [ J_n, J_{m+1} ],
\label{eq-ubjacobi}
\end{equation}
so if $[J_m,J_n]=0$ and $[J_{m+1},J_n]=0$, this then implies 
$[ J_m, J_{n+1} ] = 0$.
We remark that an operator similar to our up-boost operator 
$Y$, which we constructed in analogy to the boost in the 
Heisenberg model, has been found previously by Wadati, Hikami 
and Ujino in the context of an investigation of the systems 
with algebraic potential $V_0$ \cite{wadati}.

Finally, we note another interesting property of these integrals
of motion: Let $\Psi_0$ denote the ground state of the model, then
it has been shown in Ref.\ \cite{sriram} that 
$\sum_{j} L^{SS}_{ij} \Psi_0 = 0$ for all $i=1, \ldots, N$.
Therefore, we see that
\begin{equation}
\Psi_0^{\dagger} J_n \Psi_0 = 0
\label{eq-jn0}
\end{equation}
for all $n$. 
Thus all the $J_n$'s somehow know about the ground state and
subtract the appropriate expectation values, e.g., the
ground state expectation value of the Hamiltonian is just the 
above constant $\Phi^2 \lambda^2 N (N^2-1)/3$.

%
%

\subsection{Perturbation theory in the Lax matrices}

Looking at Eq.\ (\ref{eq-i}), we see that each $I_n$, $n>1$ 
in fact contains various powers of $I_1$.
Furthermore, in the thermodynamic limit, the $I_n$'s are not 
extensive quantities.
Thus the situation seems to be similar to
the usual problem of {\em connected} and {\em disconnected} pieces
of diagrams encountered in perturbation theory. In brief, CMR's $I_n$
seems to contain disconnected pieces and we hope that by a linked
cluster expansion, we can write new integrals of motion with
connected graphs only.

Let us be specific:
With the help of the fermionic coherent path integral \cite{ne}, 
we may rewrite the determinant 
\begin{eqnarray}
\Delta(\beta)
&= &\det[1-\beta L ], \\
&= &\int \prod_{a} dc^*_a dc_a 
        \exp[ - \sum_{jk} c^*_j [ \delta_{jk} - \beta L_{jk} ] c_k ], \\
&= &\int \prod_{a} dc^*_a dc_a 
        \exp[ - \beta \sum_{jk} c^*_j (\delta_{jk}/\beta) c_k -  c^*_j L_{jk} c_k ],
\label{eq-}
\end{eqnarray}
where $c^*_a, c_a, a=1,\ldots,N$ are Grassmann variables. Note first that
we may write this expression both for a classical $L$ {\em and} a
quantum $L$. The fact that the elements of a quantum matrix will not
necessarily commute with each other is been taken care of by the Grassmann
nature of the integration: each momentum $p_i$ will only encounter indices
$j\neq i$, otherwise the integration measure will have expressions like
$c_i c_i$ or $c^*_i c^*_i$ which are zero.

When we now include a dummy time dependence for the Grassmann variables,
i.e.\ $c^{(*)}_a=c^{(*)}_a(t)$, we can write
\begin{eqnarray}
\Delta(\beta)
&= &\int \prod_{a} dc^*_a(\tau) dc_a(\tau) 
        \exp[ - \int_{0}^{\beta} dt 
        ( 
                \sum_{jk} c^*_j(t) (\delta_{jk}/\beta) c_k(t) 
                -  c^*_j(t) L_{jk} c_k(t)
        ) ], \label{eq-Z} \\
&= & \Delta_0 \langle 
        \exp[ - \int_{0}^{\beta} dt 
        ( 
                \sum_{jk} -  c^*_j(t) L_{jk} c_k(t)
        ) ]
        \rangle_0, 
\label{eq-Z0}
\end{eqnarray}
where the average is defined as
\begin{eqnarray}
\lefteqn{
\langle F( c^*_a(t_i) c^*_b(t_j) \ldots c_g(t_k) c_h(t_l)\ldots ) \rangle_0}
& & \nonumber \\
&= &\frac{1}{\Delta_0}
  \int \prod_{a'} dc^*_{a'}(\tau) dc_{a'}(\tau) 
        \exp[ - \int_{0}^{\beta} dt 
                \sum_{j'} c^*_{j'}(t) (1/\beta) c_{j'}(t) 
        ] 
        \times \nonumber \\ & & \mbox{ }
        F( c^*_a(t_i) c^*_b(t_j) \ldots c_g(t_k) c_h(t_l)\ldots ).
\label{eq-avg}
\end{eqnarray}
This is very much like a path integral description of a many-body
partition function $Z$. 
We further note that the interaction part
$V=\sum_{jk} c^*_j(t) L_{jk} c_k(t)$ 
is just the super Lax operator $\cal L$ of SS.

The perturbation expansion is obtained by expanding Eq.\ (\ref{eq-Z})
in a power series
\begin{eqnarray}
\Delta(\beta)/\Delta_0 
&= & \sum_{n=0}^{\infty} \frac{(-1)^n}{n!} \int_{0}^{\beta}
        dt_1 dt_2\ldots dt_n
        \nonumber \\
& & \mbox{ }
        \langle
         \sum_{i_1 j_1} c^*_{i_1}(t_1) L_{i_1 j_1} c_{j_1}(t_1) \cdots
         \sum_{i_n j_n} c^*_{i_n}(t_n) L_{i_n j_n} c_{j_n}(t_n)
        \rangle_0, \label{eq-pert} \\
&\equiv & \sum_{n=0}^{\infty} \frac{(-1)^n}{n!} \Delta_n, 
\label{eq-inpert}
\end{eqnarray}
and $\Delta_n\sim\beta^n I_n$.
The last equation is obtained by comparison with Eq.\ (\ref{eq-in})
and $\Delta_0=1$. 
Note that $I_n =0$ in Eq.\ (\ref{eq-inpert}) for all $n>N$. E.g.\ for
$N=2$, we have
$I_3 \sim \sum_{i_1 j_1 i_2 j_2}^{N=2} \sum_{i_3 j_3}^{N=2}
        L_{i_1 j_1} L_{i_2 j_2} L_{i_3 j_3}
        \langle c^*_{i_1} c_{j_1} c^*_{i_2} c_{j_2} c^*_{i_3} c_{j_3}
        \rangle_0
$ and clearly $i_3, j_3$ always take index values already covered by 
$\{ i_1, j_1, i_2, j_2 \}$. Thus the bracket $\langle\rangle$ is zero
by the Grassmann character of the $c$'s.

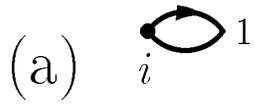
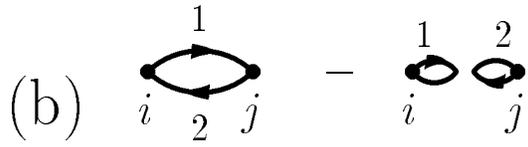
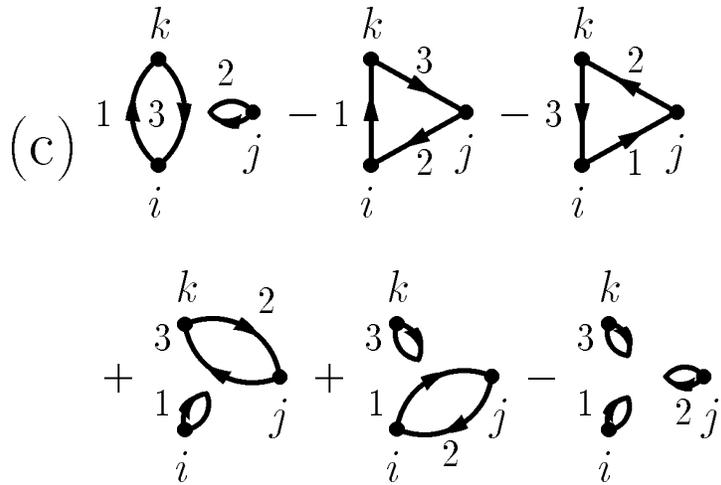
\begin{figure}
  \caption{
   Diagrammatic representation of Eq.\ (\protect\ref{eq-d}).
   Each line is labeled by a number indicating precedence in
   the corresponding matrix product. $1$ corresponds to the
   right-most matrix.
  }
\label{fig-d13}
\end{figure}

Let us now calculate the first few orders of $\Delta(\beta)$. 
With $g_i$ being a dummy propagator, we find
\begin{mathletters}
\label{eq-d}
\begin{eqnarray}
\Delta_1
&= & -\beta \sum_{i} L_{ii} g_i
\label{eq-d1} \\
\Delta_2
&= & \frac{1}{2} \beta^2 
\sum_{ij} \left( L_{ij} L_{ji} - L_{ii} L_{jj} \right) g_i g_j
\label{eq-d2} \\
\Delta_3
&= & -\frac{1}{3!} \beta^3
        \sum_{\stackrel{{\scriptstyle i_1,i_2,i_3,}}{j_1,j_2,j_3}}
        L_{i_1 j_1} L_{i_2 j_2} L_{i_3 j_3}
        \langle c^*_{i_1} c^*_{i_2} c^*_{i_3} c_{j_1} c_{j_2} c_{j_3}
        \rangle_0 \nonumber \\
&= & -\frac{1}{3!} \beta^3
        \sum_{ijk} \left(
        L_{ik} L_{jj} L_{ki} 
        - L_{ik} L_{ji} L_{kj}
        - L_{ij} L_{jk} L_{ki} \right. \nonumber \\ 
& & \left. \mbox{ }
        + L_{ii} L_{jk} L_{kj}
        + L_{ij} L_{ji} L_{kk}
        - L_{ii} L_{jj} L_{kk}
        \right) g_i g_j g_k.
\label{eq-d3} 
\end{eqnarray}
\end{mathletters}
Introducing the diagrammatic notation
$i\, \bullet\!\!\!\longrightarrow j \equiv L_{ij}$,
we can represent these expressions by their graphs as in
Fig.\ \ref{fig-d13}.
Note that only in Eq.\ (\ref{eq-d3}) do we need to worry about the 
ordering of the matrix products. If we ignore that
ordering for the moment --- the classical case --- we have
\begin{eqnarray}
\Delta_3
&= & -\frac{1}{3!} \beta^3
\sum_{ijk} 
 \left[ 
 3 L_{ii} L_{jk} L_{kj} 
 -2 L_{ij} L_{jk} L_{ki}
 - L_{ii} L_{jj} L_{kk} 
 \right] g_i g_j g_k. 
\label{eq-d3c}
\end{eqnarray}
We see that the second term in Eq.\ (\ref{eq-d3c}), representing the 
fully connected diagram, is actually just 
$\mbox{const.} \times \mbox{Tr} (L^3)$.

Let us see what these expressions tell us:
(i) We see that in fact the $\Delta_n$'s are just the $I_n$'s of CMR
with $g_i=1$.
(ii) We observe that the fully connected diagrams give the traces of 
powers of $L$ in the classical case as was expected from the well-known
matrix formula $\ln \det A = \mbox{Tr} \ln A$. Thus these connected diagrams
are the quantum analogue of the {\em classical} integrals of motion.
(iii) The ordering of the matrix products becomes important for $n>2$,
thus necessitating order labeling of diagrams.

%
%

\subsection{Constructing connected diagrams}

We now want to rewrite the perturbation expansion (\ref{eq-pert})
such that we only use fully connected diagrams. And we want to do
this such that we can minimise the ordering problems coming from the
quantum character of the Lax matrix.
The basic program is due to Thiele and know as the linked cluster
theorem. It can be summarised as follows:
We resum the series (\ref{eq-inpert}) as
\begin{equation}
\Delta(\beta) = \exp[ -\sum_{n=1} \frac{\beta^n}{(n-1)!} T_n ],
\label{eq-tnpert}
\end{equation}
and use it to define the $T_n$'s. Comparing Eq.\ (\ref{eq-inpert})
and (\ref{eq-tnpert}), we find up to $n=4$,
\begin{mathletters}
\label{eq-t}
\begin{eqnarray}
T_1 
&= & I_1, \\
&= &\sum_{i} p_i, \label{eq-t1} \\
T_2 
&= & I_1^2 - 2 I_2, \\
&= &
\sum_i p_i^2 + \sum_{ij}' \alpha_{ij}^2, 
\label{eq-t2} \\
T_3 
&= & I_1^3 - 3 I_1 I_2 + 3 I_3, \\
&= &
\sum_i p_i^3 + 3 \sum_{ij}' \alpha_{ij}^2 p_i 
\label{eq-t3} \\
T_4
&= & I_1^4 - 4 I_2 I_1^2 +2 I_2^2 + 4 I_3 I_1 - 4 I_4, \\ 
&= & \sum_i p_i^4
+2 \sum_{ij}' \alpha_{ij}^2 \left[ p_j^2 + p_i p_j + p_i^2 \right] 
+ \mbox{Tr} \alpha^4
\nonumber \\ & &
-2 \sum_{ij}' \alpha''_{ij} \alpha_{ij}
-2 \sum_{ij}' (\alpha'_{ij})^2
-4 i \sum_{ij}' \alpha'_{ij} \alpha_{ij} p_i.
\label{eq-t4}
\end{eqnarray}
\end{mathletters}
Since CMR have already proven $[I_n,I_m]=0$, there is no ordering 
problem for the $I_n$'s in the construction of the $T_n$'s and
we furthermore have $[T_n,T_m]=0$.
Since $T_2$ is, up to a constant, the Hamiltonian, this implies
both involution and invariance.

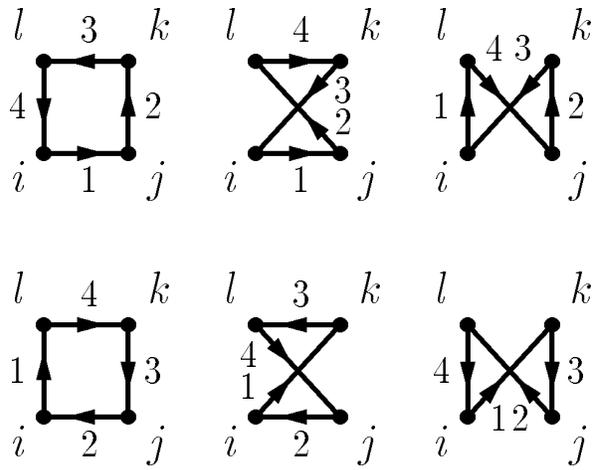
\begin{figure}
  \caption{
   The $6$ diagrams associated with $T_4$.
  }
\label{fig-t4}
\end{figure}

As expected, we find that each $T_n$ corresponds to the fully 
connected diagrams of the series (\ref{eq-inpert}). 
We can now directly use the diagrammatic approach to 
construct the $T_n$'s. However, for a given $n$, there are $(n-1)!$
different labeled diagrams and thus different matrix orderings.
Each diagram itself is an ordered operator expression and it is
quite tedious to get them into a form as in Eq.\ (\ref{eq-t})
with all momenta to the right.
As an example, we give the diagrams for $T_4$ in 
Fig.\ \ref{fig-t4}.

Ignoring matrix and quantum ordering, the resultant expressions
for the $T_n$'s are equal to the classically invariant
traces of Eq.\ (\ref{eq-ct}). 
Thus we may hope that due to the special form of the quantum Lax 
matrix $L$, the matrix product order somehow is unimportant
and the $T_n$'s are just the quantum traces 
$\mbox{Tr} L^n$.
Explicitly calculating the quantum traces, we find that indeed
up to $n=3$, we have $T_n = \mbox{Tr} L^n$.
However, for $n=4$, the quantum trace includes the nonzero term
$-2 \sum_{ijk}' \alpha_{ij} \alpha_{jk} \alpha'_{ki}$.
Note that this term has a factor
$[ \lambda(\lambda-1) ]^{2/3}$ \cite{missing}.
But as shown in section
\ref{sec-qci}, such a term does not arise in the $I_n$'s and
consequently also not in the $T_n$'s.
Therefore, the $T_n$'s are not simply the quantum traces.

Note that these expressions are again very close to the ones 
obtained for $J_n$. 
However, as before in the classical case, we see that 
already for $n=2,3$, there are the same constants
in the $J$ expressions which do not appear in the $T$ expressions.

%
%
\subsection{Relation between invariants}
\label{sec-rbi}

We again would like to see if we can express Shastry's integrals
in terms of our $T_n$'s. As we have seen in section \ref{sec-crbi}
for the classical case, we expect this relation to be algebraic.
Fortunately, as shown in the last section, fractional powers
of $\lambda$ neither appear in the $J_n$'s nor in the $T_n$'s
so that no a priori reasons forbid an algebraic relationship in
the quantum case. 
         
Furthermore, given two sets of integrals of motion $\{T_n\}$ and
$\{J_n\}$, we know that commutators of integrals are themselves
integrals of motion, and since asymptotically these integrals
evaluate to zero, the two sets of integrals can be 
simultaneously diagonalized.
A relationship between asymptotic integrals of the form
${\cal J}_n= A_n[ \{{\cal T}_m\} ]$ can always be found, since
either set of integrals gives an algebraically complete
set of symmetric polynomials of increasing degree.
Suppose we have such a relationship. Then, the operators
$J_n$ and $A_n[ \{T_m\} ]$ have the same eigenvalues in the same
basis, hence must be the same operator, and so there must
exist a relationship
$J_n = A_n[ \{T_m\} ]$
between the operators themselves.
 
Replacing $\alpha_{ij}$ and $\chi_{ij}$ by their appropriate
definitions, using the explicit form for the derivatives as given 
in the appendix and counting powers of $\lambda$ and $p$'s, we
then have
\begin{mathletters}
\label{eq-jt}
\begin{eqnarray}
J_1 &= &T_1, 
\label{eq-jt1} \\
J_2 &= &
\Phi^2 \lambda [3 + \lambda(N-2)] N (N-1)/3 + T_2, 
\label{eq-jt2} \\
J_3 &= &
\Phi^2 \lambda [3 + \lambda(N-2)] (N-1) T_1 + T_3, 
\label{eq-jt3} \\
J_4 &= &
 \left[
  -75 + 120\lambda - 48\lambda^2 + (55 - 110\lambda + 52\lambda^2) N
  + ( -5 + 25\lambda - 18\lambda^2) N^2 + 2 \lambda^2 N^3
 \right] \times
\nonumber \\ 
& & \mbox{ } \lambda^2 N(N-1) \Phi^4/15 + 
   \left[2 + \lambda(N-2)\right] \lambda\Phi^2 T_1^2 + 
\nonumber \\ 
& & \left[10 - 12\lambda + 11 (\lambda-1) N + (2\lambda-1) N^2\right]
   \lambda\Phi^2 T_2 + T_4.
\label{eq-jt4}
\end{eqnarray}
\end{mathletters}
Hence we have succeeded in writing Shastry's quantum integrals
in terms of the $T_n$'s which in turn are derived from Calogero's
quantum integrals for up to $n=4$. Again, as in the classical
case, this relationship is not linear, since we observe the
$T_1^2$ term in Eq.\ (\ref{eq-jt4}).
And only if we restrict ourselves to the potential $V_0(x)= g^2/x^2$
by taking the limit $\Phi\rightarrow 0$, do we find that both sets 
of integrals of motion are identical. 

With $\Psi_0$ the $N$-particle ground state as before, we may
use Eq.\ (\ref{eq-jn0}) and hence relate the expectation values
of various $T_n$'s. 
E.g., 
$\Psi_0^{\dagger} T_2 \Psi_0 =
\Phi^2 \lambda [3 + \lambda(N-2)] N (N-1)/3$ 
and
$\Psi_0^{\dagger} T_3 \Psi_0 =
-\Phi^2 \lambda [3 + \lambda(N-2)] (N-1) \Psi_0^{\dagger} T_1 \Psi_0 
\sim \Psi_0^{\dagger} P \Psi_0 = 0$.
We further note that as in the classical case, 
we may define 
new non trivial constants of motion $K_n= J_n-T_n$ for $\Phi\neq 0$.
We then have
$\Psi_0^{\dagger} K_n \Psi_0 = - \Psi_0^{\dagger} T_n \Psi_0$.
Unfortunately, we can not give a simple formula directly in
terms of the Lax matrices for the construction of the $K_n$'s
analogous to Eq.\ (\ref{eq-kn}).

%
%

\section{The asymptotic limit}
\label{sec-limit}

In the asymptotic $t \rightarrow\infty $ limit --- equivalent
to the low-density limit --- the elements of the Lax matrix
are no longer operators, but numbers. Thus explicit calculations
are much easier and we hope that we can study the connection
between asymptotic Calogero and Shastry integrals of motion
in more detail than in the last section.

\subsection{Asymptotics of Calogero's integrals}

The asymptotic form for the Lax matrix L gives a corresponding
asymptotic form for the Calogero integrals 
$I_n\rightarrow {\cal I}_n$.
We define a generating function of the asymptotic Calogero
integrals,
\begin{equation}
{\cal D}[z,\lambda] 
= \frac{1}{2} \left[
 \prod_{j=1}^{N} (1 - z p_j - i\lambda z)
 +\prod_{j=1}^{N} (1 - z p_j + i\lambda z) 
 \right],
\end{equation}
and then have
\begin{equation}
\det [1 - z L^\infty] = {\cal D}[z,\sqrt{\lambda(\lambda-1)}].
\end{equation}
We also define the elementary symmetric functions of the $N$
variables $p_j$ as
\begin{equation}
a_r[p] = \sum_{1\leq j_1<\ldots<j_r}^{N} 
 p_{j_1} \cdots p_{j_r},
\end{equation}
and for convenience $a_0=1$, and $a_{-r}=0$.
Then
\begin{equation}
{\cal D}[z,\lambda] = \sum_{r=0}^{\infty} 
(-\lambda^2 z^2)^r a_{N-2r}[1-zp].
\end{equation}
In particular,
\begin{equation}
 {\cal I}_N = a_N[p] - \lambda(\lambda-1) a_{N-2}[p] +
 \lambda^2(\lambda-1)^2 a_{N-4}[p] - \ldots,
\end{equation}
and the other integrals can be constructed via
\begin{equation}
\left[ \sum_{k=1}^{N} \frac{\partial}{\partial p_k}, {\cal I}_j \right]
 = (N-j+1) {\cal I}_{j-1}.
\end{equation}
Since the elementary symmetric functions $a_j$ obey the same relationship, 
this also gives us the expression for ${\cal I}_j$ as a linear
combination of $a_j$, $a_{j-2}$, $a_{j-4}$, $\ldots$.
The general expression for ${\cal I}_j$ in terms of the
$a_r[p]$ can be obtained using 
\begin{equation}
a_r[ 1-zp ] =
 \sum_{j=0}^{N} (-z)^{r-j}
  \left( \begin{array}{c} N+j-r \\ j \end{array} \right)
  a_{r-j}[p],
\end{equation}
and we find
\begin{equation}
{\cal I}_j =
 \sum_{j=0}^{N} [-\lambda(\lambda-1)]^r
  \left( \begin{array}{c} N+2r-j \\ 2r \end{array} \right)
  a_{j-2r}[p].
\end{equation}

\subsection{Some generating functions}

Let us define the quantity
\begin{equation}
{\cal Z}[z,\lambda] =
 \prod_{j=1}^{N}
  [ 1 - z(p_j+i\lambda) ]
 \equiv {\cal D}[z,\lambda] - i \lambda z {\cal N}[z,\lambda].
\end{equation}
The symmetric part of ${\cal Z}$ is
${\cal D}= ( {\cal Z}[z,\lambda]+{\cal Z}[z,-\lambda])/2$
as in the previous section, while the antisymmetric part
$( {\cal Z}[z,\lambda]-{\cal Z}[z,-\lambda])/2$ 
is given by
\begin{equation}
 i \lambda z {\cal N}[z,\lambda]
  = i \lambda z
   \sum_{r=0}^\infty
   (-\lambda^2 z^2)^r a_{N-2r-1}[1 - z p].
\end{equation}
For $z$ real, these are the real and imaginary parts of ${\cal Z}$,
respectively. This expression for ${\cal Z}$ is the standard generating 
function for  the elementary symmetric functions, so that
\begin{equation}
{\cal Z}[z,\lambda] = \sum_{j=0}^N (-z)^j a_j[p + i \lambda ].
\end{equation}
Clearly, then, we have ${\cal I}_j= 
\mbox{Re}\{a_j[p+ i\sqrt{\lambda(\lambda-1)}] \}$.
Taking the logarithm of ${\cal Z}$,
\begin{equation}
\ln {\cal Z}[z,\lambda] =
 \sum_{j=1}^N \ln[ 1-z(p+j+i \lambda)],
\end{equation}
one advantage of using the generating function ${\cal Z}$ is clear when
one anticipates the thermodynamic limit.
We now consider
\begin{equation}
\frac{\partial}{\partial z} \ln {\cal Z}
 = \sum_{j=1}^{N}
   -\frac{ p_j + i \lambda }
        {1 - z( p_j+ i \lambda)}.
\end{equation}
This, however, is the generating function for the symmetric power
sums
\begin{equation}
b_r[p] = \sum_{j=1}^N p_j^r
\end{equation}
since
\begin{equation}
P[z] = \sum_{r=0}^\infty z^r b_{r+1}[p]
     = \sum_{j=1}^N \frac{p_j}{1-z p_j}
\end{equation}
and we see that
\begin{equation}
\frac{\partial}{\partial z} \ln {\cal Z}[z,\lambda]
 = \sum_{r=0}^{\infty}
    z^r b_{r+1}[p + i \lambda]
 \equiv P[z,\lambda].
\end{equation}

\subsection{Asymptotics of Shastry's integrals}

Shastry's integrals also approach an asymptotic form
$J_j \rightarrow {\cal J}_j$, and similarly we define an 
asymptotic generating function
\begin{equation}
{\cal G}[z,\lambda]\equiv \frac{{\cal N}[z,\lambda]}{{\cal D}[z,\lambda]}.
\end{equation}
then, we see that
${\cal N}[z,\lambda] = {\cal G}[z,\lambda] {\cal D}[z,\lambda],$
so expanding, we have
\begin{equation}
 \sum_{k=1}^N (-z)^k \mbox{Im}\{ a_k[p + i \lambda] \}
  = \lambda \sum_{k=1}^\infty \sum_{j=1}^\infty
   (-1)^{j-1} {\cal J}_{j-1} 
   \mbox{Re}\{a_{k-j}[p + i \lambda] \}.
\end{equation}
Equating powers of $z$, this gives
\begin{equation}
 \mbox{Im}\{ a_k[p + i \lambda] \}
  = \lambda \sum_{j=1}^k
   (-1)^{j-1} {\cal J}_{j-1} 
   \mbox{Re}\{a_{k-j}[p + i \lambda] \}.
\end{equation} 
More explicitly,
\begin{equation}
\lambda^{-1} \mbox{Im}\{ a_k[p + i \lambda] \} =
 N\mbox{Re}\{ a_k[p+i\lambda]\}
 - {\cal J}_1 \mbox{Re}\{a_{k-1}[p+i\lambda]\}
 + \ldots +
 (-1)^k {\cal J}_k.
\end{equation}
This allows us to iterate and find ${\cal J}_k$ in terms of
the other ${\cal J}_j$, 
\begin{equation}
 {\cal J}_k =
  {\cal J}_{k-1} \mbox{Re}\{a_1[p+i\lambda]\}
  - \ldots -
  (-1)^k N \mbox{Re}\{ a_k[p+i\lambda]\}
  + (-1)^k \lambda^{-1} \mbox{Im}\{ a_{k+1}[p+i\lambda]\}.
\end{equation}

Since ${\cal I}_j = \mbox{Re}\{a_j[p+i\lambda]\}$, this
recursion relation also relates the asymptotic forms of
the CMR and SS integrals. However, this relation is between
asymptotic integrals with the same parameter, and such integrals
do not even commute.

%
%
\section{Conclusions}
\label{sec-concl}

Our original intent and hope at taking up the present work was
to give a simple connection between the integrals of motion of
CMR and the recently discovered integrals of SS.
In fact, we were speculating that due to the special structure
of the quantum Lax matrix, we would simply find
$T_n \sim J_n$.

Quite the opposite has happened:
In the quantum case, though we have succeeded in rewriting the
$I_n$
integrals of CMR into extensive quantities $T_n$, we have however
failed to give a simple formula for the connection of these
$T_n$'s to the $J_n$ integrals of SS for all but the simplest
potential $V_0(x)= g^2/x^2$.
In general, we do find a complicated algebraic relationship which 
gives rise to yet another set of non trivial integrals of motion
$K_n$.

In the classical case, we show that the quantum definition of SS
may also be used to construct classical integrals of motion.
Hence here the situation now is just as in the quantum case and
again we show that we may reexpress the integrals of SS in
terms of the classical integrals of Moser.
Again, the difference of these two sets of integrals 
vanishes only for the potential $V_0$ and otherwise may be 
used to define new constants and this time, we can give an 
explicit formula for the $K_n$'s directly in terms on the
Lax $L$ matrix.
Indeed, we may even define a one-parameter family of integrals
of motion in the classical case, e.g.,
\begin{equation}
R_n(\delta) = \mbox{Tr} \left[ L^n (1 + \delta \Delta') \right],
\end{equation}
with $\Delta'_{ij}=1$ if $i\neq j$ and zero otherwise.
Then $R_n$ interpolates between Moser's integrals $R_n(0)= T_n$ 
and the integrals of SS $R_n(1)= J_n$.

Most of the previous formulas are given in terms of 
$\alpha$'s and $\chi$'s and are thus valid not only for the
hyperbolic pair potential $V_h$, but also for the trigonometric
$V_t$ and the rational $V_0$ after taking the appropriate limits 
as mentioned in the beginning of section \ref{sec-classical}. 
However, we have found an up-boost only for the classical and
the quantum many-body system with algebraic potential 
$V_0(r)= g^2/r^2$.  
Also, the elliptic potential $V_e = 1/\mbox{sn}^2$, is not
included in this study: Although the classical integrals of
Moser and the quantum integrals of CMR are valid for this
potential, the proof of SS does no longer hold both for the
classical and the quantum case. This is due to the fact that
the row and column sums of the elliptic Lax $M$ matrix are
no longer zero.
The Ansatz $J'_n= \mbox{Tr} L^n g[\{x_j\}]$ gives an equation
for the coordinate dependent matrix $g$ as
$\frac{\partial}{\partial t} g = Mg - gM$.
Unfortunately, we have not found a non trivial such $g$ 
such that it gives the SS integrals of motion in the 
elliptic case. 

\acknowledgments
R.A.R.\ gratefully acknowledges financial support from the 
Alexander von Humboldt foundation. 

%
%
\appendix

\section{Derivatives}

We have defined $\gamma_{ij}= \coth [\Phi(x_i-x_j)]$ and so
$\chi_{ij} = \Phi \lambda \gamma_{ij}$ and
$\alpha_{ij}= \Phi \sqrt{\lambda(\lambda-1)} \gamma_{ij}$.
Then the derivatives are
\begin{mathletters}
\begin{eqnarray}
\frac{\partial \chi_{ij}}{\partial x_i}= \chi_{ij}'
&= & \Phi^2 \lambda [1 - \gamma_{ij}^2] \\
\chi_{ij}''
&= & -2 \Phi^3 \lambda \gamma_{ij} (1 - \gamma_{ij}^2) \\
\chi_{ij}'''
&= & -2\Phi^4\lambda( 1 - 4\gamma_{ij}^2 + 3 \gamma_{ij}^4).
\end{eqnarray}
\end{mathletters}
The same relations hold for $\alpha_{ij}$, i.e.\
$
\alpha_{ij}'
=  \Phi^2 \sqrt{\lambda(\lambda-1)} [1 - \gamma_{ij}^2] $, $
\alpha_{ij}''
=  -2 \Phi^3 \sqrt{\lambda(\lambda-1)} \gamma_{ij} (1 - \gamma_{ij}^2) 
$, and $
\alpha_{ij}'''
=  -2\Phi^4\sqrt{\lambda(\lambda-1)}( 1 - 4\gamma_{ij}^2 + 3 \gamma_{ij}^4),
$
and, lastly,
$
\gamma_{ij}'
=  \Phi [1 - \gamma_{ij}^2] $, $
\gamma_{ij}''
=  -2 \Phi\gamma_{ij} (1 - \gamma_{ij}^2) 
$, and $
\gamma_{ij}'''
=  -2\Phi( 1 - 4\gamma_{ij}^2 + 3 \gamma_{ij}^4).
$

\begin{references}
\bibitem{asym} We would still need to prove that, 
        for $t\rightarrow\pm\infty$, $L$ takes the form
        of $L^{\infty}$, at least in the sense of weak
        convergence. However, we have explicitly checked
        the commutation relations of the $I_n$'s for up to
        $N=5$ particles and $n=5$.
\bibitem{missing} 
        This term seems to be missing in Ref.\ \cite{opq}.
\end{references}
\end{document}